\documentclass[aip,apl,reprint,final]{revtex4-2}

\pdfoutput=1  % include for ArXiv

\usepackage[colorlinks=true,allcolors=blue]{hyperref} % remove for APL
\usepackage{graphicx}% Include figure files
\usepackage{amsmath,amssymb,mathrsfs,array,longtable,delarray}
\usepackage{latexsym}
\usepackage{dcolumn}% Align table columns on decimal point
\usepackage{bm}% bold math
\usepackage{textcomp} % for upright greek letters  {\textmu} {\textohm}

\begin{document}

\title{Non-adiabatic single-electron pumps in a dopant-free GaAs/AlGaAs 2DEG}

\author{B. Buonacorsi}
\affiliation{Institute for Quantum Computing, University of Waterloo, Waterloo N2L 3G1, Canada}
\affiliation{Department of Physics, University of Waterloo, Waterloo N2L 3G1, Canada}
\author{F. Sfigakis}
\altaffiliation{corresponding author: francois.sfigakis@uwaterloo.ca}
\affiliation{Institute for Quantum Computing, University of Waterloo, Waterloo N2L 3G1, Canada}
\affiliation{Northern Quantum Lights Inc., Waterloo N2B 1N5, Canada}
\affiliation{Department of Chemistry, University of Waterloo, Waterloo N2L 3G1, Canada}
\author{A. Shetty}
\affiliation{Institute for Quantum Computing, University of Waterloo, Waterloo N2L 3G1, Canada}
\affiliation{Department of Chemistry, University of Waterloo, Waterloo N2L 3G1, Canada}
\author{M. C. Tam}
\author{H. S. Kim}
\affiliation{Department of Electrical and Computer Engineering, University of Waterloo, Waterloo N2L 3G1, Canada}
\affiliation{Waterloo Institute for Nanotechnology, University of Waterloo, Waterloo N2L 3G1, Canada}
\author{S. R. Harrigan}
\affiliation{Institute for Quantum Computing, University of Waterloo, Waterloo N2L 3G1, Canada}
\affiliation{Department of Physics, University of Waterloo, Waterloo N2L 3G1, Canada}
\affiliation{Waterloo Institute for Nanotechnology, University of Waterloo, Waterloo N2L 3G1, Canada}
\author{\\ F. Hohls}
\affiliation{Physikalisch-Technische Bundesanstalt (PTB), 38116 Braunschweig, Germany}
\author{M. E. Reimer}
\affiliation{Institute for Quantum Computing, University of Waterloo, Waterloo N2L 3G1, Canada}
\affiliation{Department of Physics, University of Waterloo, Waterloo N2L 3G1, Canada}
\affiliation{Northern Quantum Lights Inc., Waterloo N2B 1N5, Canada}
\affiliation{Department of Electrical and Computer Engineering, University of Waterloo, Waterloo N2L 3G1, Canada}
\author{Z. R. Wasilewski}
\affiliation{Institute for Quantum Computing, University of Waterloo, Waterloo N2L 3G1, Canada}
\affiliation{Department of Physics, University of Waterloo, Waterloo N2L 3G1, Canada}
\affiliation{Northern Quantum Lights Inc., Waterloo N2B 1N5, Canada}
\affiliation{Department of Electrical and Computer Engineering, University of Waterloo, Waterloo N2L 3G1, Canada}
\affiliation{Waterloo Institute for Nanotechnology, University of Waterloo, Waterloo N2L 3G1, Canada}
\author{J. Baugh}
\altaffiliation{baugh@uwaterloo.ca}
\affiliation{Institute for Quantum Computing, University of Waterloo, Waterloo N2L 3G1, Canada}
\affiliation{Department of Physics, University of Waterloo, Waterloo N2L 3G1, Canada}
\affiliation{Northern Quantum Lights Inc., Waterloo N2B 1N5, Canada}
\affiliation{Department of Chemistry, University of Waterloo, Waterloo N2L 3G1, Canada}
\affiliation{Waterloo Institute for Nanotechnology, University of Waterloo, Waterloo N2L 3G1, Canada}

\begin{abstract}
We have realized quantized charge pumping using non-adiabatic single-electron pumps in dopant-free GaAs two-dimensional electron gases (2DEGs). The dopant-free III-V platform allows for ambipolar devices, such as p-i-n junctions, that could be combined with such pumps to form electrically-driven single photon sources. Our pumps operate at up to 0.95 GHz and achieve remarkable performance considering the relaxed experimental conditions: one-gate pumping in zero magnetic field and temperatures up to 5 K, driven by a simple RF sine waveform. Fitting to a universal decay cascade model yields values for the figure of merit $\delta$ that compare favorably to reported modulation-doped GaAs pumps operating under similar conditions. The devices reported here are already suitable for optoelectronics applications, and with further improvement could offer a route to a current standard that does not require sub-Kelvin temperatures and high magnetic fields.
\end{abstract}

\maketitle

At low temperatures, non-adiabatic single electron pumps show quantized current plateaus $I_{\text{pump}}=nef$,\cite{Blumenthal07,FujiwaraA08,Kaestner08A,Kaestner08B} where $e$ is the electron charge, $f$ is the radio frequency (RF) of an ac signal applied to a local gate, and $n$ is the number of electrons pumped from source to drain during each RF cycle. These pumps have been the subject of several reviews\cite{Pekola13,Janssen14,Kaestner15,Kaneko16,Giblin19,Laucht21} for applications in quantum metrology, and have been variously referred to as tunable-barrier pumps, one-parameter pumps, ratchet pumps, or dynamic quantum dots. They do not require an applied source-drain bias to drive current, operate at high frequencies ($\sim$\,GHz), and are distinct from adiabatic turnstile pumps,\cite{Geerligs90,Kouwenhoven91-A} which require a finite source-drain bias and a minimum of two RF gates to operate (with typical $f < 20$ MHz).

The most commonly used pump architectures involve modulation-doped GaAs/AlGaAs two-dimensional electron gases (2DEG),\cite{Giblin12,Bae15,Stein15} because of their simple operation (one-gate pumping), relative ease of fabrication, and high performance. The lowest relative uncertainties in metrological measurements of pumped current in GaAs were achieved at $f$\,= 545 MHz\cite{Stein15} and at $f$\,= 600 MHz\cite{Stein17} using low temperatures ($T$\,$<$\,100 mK), high magnetic fields ($B$\,$>$\,10 T), and shaped RF pulses. The ease of fabrication allows the positioning of non-invasive detectors\cite{Field93} near pumps, thus enabling the individual counting of pumped electrons and the identification of error mechanisms,\cite{Fricke13,Giblin16} and allows the integration of pumps into more complex circuits, towards a self-referenced quantum current standard.\cite{Fricke14} Non-adiabatic pumps have also been realized in other material systems: graphene,\cite{Connolly13} InAs nanowires,\cite{dHollosy15} and silicon.\cite{FujiwaraA08,Yamahata14A,Wenz16,Rossi18,Yamahata19,MurrayR18,Rossi14,MurrayR20,Yamahata16,Giblin20,ZhaoR17}
In some Si-based pumps, the dynamic quantum dot consists of an accidental charge trap located near the transport channel.\cite{Yamahata14A,Wenz16,Rossi18,Yamahata19} Among the more conventional silicon-based pumps,\cite{FujiwaraA08,MurrayR18,Rossi14,MurrayR20,Giblin20,Yamahata16,ZhaoR17} the lowest relative uncertainties in metrological measurements of pumped current were achieved at $f$\,=\,1 GHz and $T$\,=\,4 K,\cite{Giblin20} and at $f$\,=\,1 GHz and $T$\,=\,0.3 K.\cite{ZhaoR17} Remarkably, those measurements were performed in zero magnetic field, and using only a simple RF sine wave.

One possible reason for the relative improved performance (in zero magnetic field, using only a simple RF sine waveform) of some Si pumps over modulation-doped GaAs pumps may be the ability of the former to generate higher electric fields locally. Indeed, experiments that vary the confinement-potential shape in GaAs quantum dots used as non-adiabatic single electron pumps have shown promising results.\cite{SeoM14,Ahn17A} Improved pump performance relative to modulation-doped GaAs pumps could perhaps be obtained in GaAs-based dopant-free heterostructure-insulator-gate field effect transistor (HIGFET) geometry.\cite{Francois10, ChenJCH12, WangDQ13} Such accumulation mode transistors have already been used to produce Hall bars,\cite{Harrell99,Wendy10,Croxall13,Sebastian16,Croxall19,arxiv2012.14370} quantum wires,\cite{Reilly01, Klochan06, Sarkozy09-A, Srinivasan20} and quantum dots,\cite{See10, Klochan11, See12, Wendy13, Bogan17, Bogan18} with demonstrated superior performance in terms of low disorder, suppressed random telegraph switching (RTS) events/noise, and cooldown-to-cooldown (even device-to-device) reproducibility relative to their GaAs modulation-doped counterparts.\cite{Luke16}

\begin{figure}[t]
    \includegraphics[width=0.95\columnwidth]{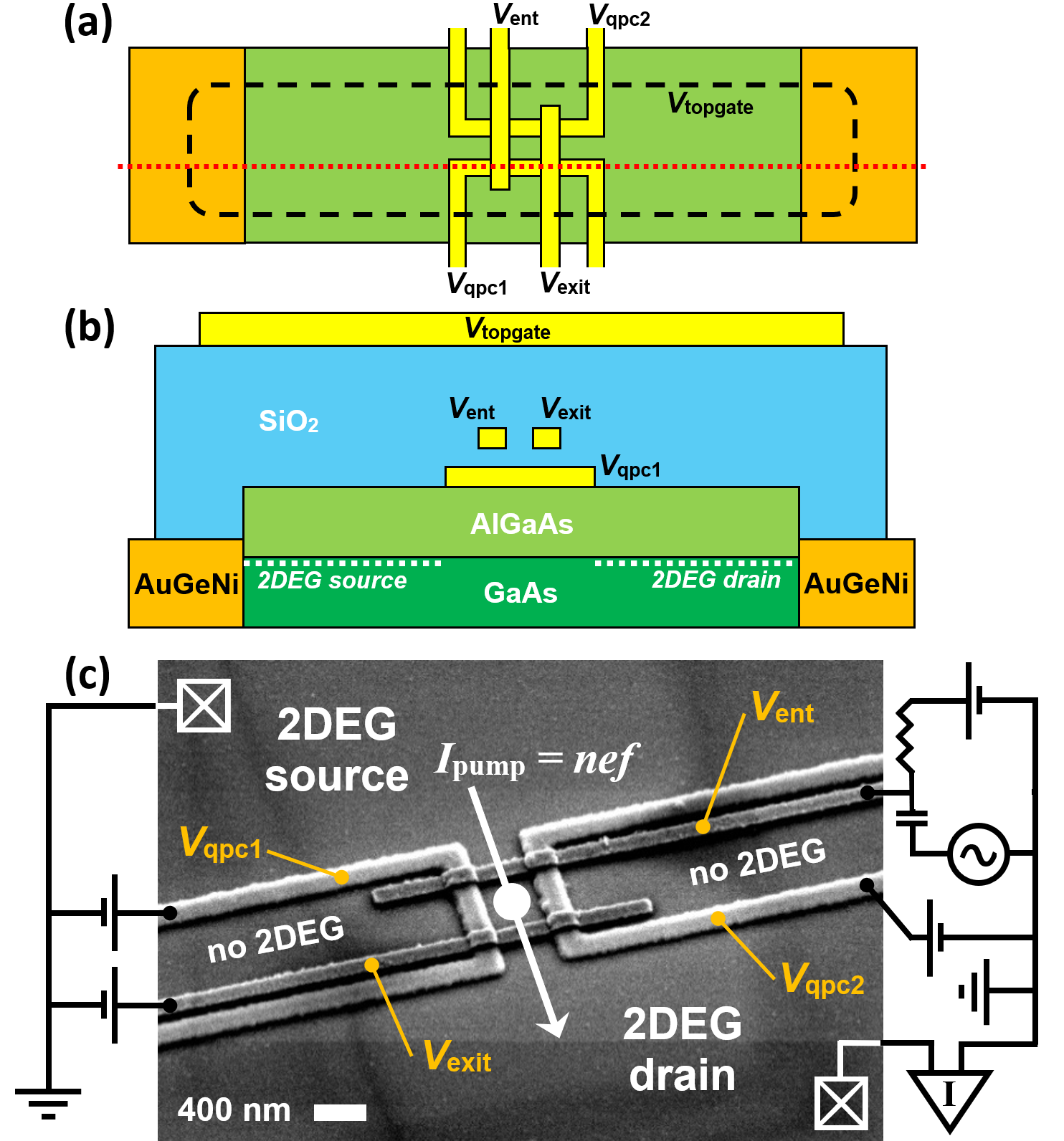}
    \caption{(a) Diagram of gate layout design A (view from above). Only the outline of $V_{\text{topgate}}$ is visible (black dashed line). (b) Cross-sectional view (sideview) of gate layout design A, along the cut indicated by the red dotted line in panel a. (c) Measurement circuit diagram and scanning electron microscope (SEM) image of a single electron pump in a dopant-free 2DEG with gate layout design A, similar to device A1. All fine gates ($V_{\text{qpc1}}, V_{\text{qpc2}}, V_{\text{exit}}$, and $V_{\text{ent}}$) were written by electron beam lithography. There are three levels of gates, each separated by a dielectric. A one-dimensional channel, defined electrostatically by gates $V_{\text{qpc1}}$ and $V_{\text{qpc2}}$, is connected at both ends to a 2DEG (source and drain ohmic contacts $\boxtimes$) induced by a global topgate $V_{\text{topgate}}$ deposited on top of the device (not shown in panel c). Entrance and exit barrier gates ($V_{\text{ent}}$ and $V_{\text{exit}}$) are used to form a dynamically driven quantum dot (white dot); they are isolated from $V_{\text{qpc1}}$ and $V_{\text{qpc2}}$ by 30 nm of SiO$_2$. The topgate is isolated from the recessed AuGeNi ohmic contacts and $V_{\text{qpc}}$ gates by 300 nm of SiO$_2$. By RF modulation of the entrance barrier voltage, electrons are collected from the 2DEG source and ejected into the 2DEG drain [see Fig.\,\ref{Fig-RFcycle}e], resulting in a time-averaged quantized current $I_{\text{pump}}=nef$ [see Fig.\,\ref{Fig-Inef}] measured by a DC current pre-amplifier at the 2DEG drain ($\bigtriangledown$). Note there is no bias applied between source and drain.}
    \label{Fig-SEM}
\end{figure}  % Original fig file: Fig 1 - SEM v5 hi-res.png

Since GaAs has a direct bandgap, dopant-free non-adiabatic single electron pumps could be useful for applications in quantum optoelectronics. For example, by juxtaposing a non-adiabatic single electron pump next to a ``lateral'' p-i-n junction in the same device, one could realize an all-electrical, on-demand, and high-rate single photon source.\cite{Blumenthal07} These could be manufactured using industry-standard fabrication techniques; their compactness and controllable positioning would allow them to be scaled to arrays (\textit{e.g.}, for multiplexing\cite{Laferriere20}) or integrated with other 2DEG-based devices for new functionalities (\textit{e.g.}, with spin qubits\cite{CerfontaineP20,KandelYP21,QiaoH21} for spin-to-photon conversion\cite{HsiaoTK20}). The first building block of such single photon sources is of course a lateral p-i-n junction, already realized in a dopant-free GaAs/AlGaAs system by others\cite{Dai13,Dai14,ChungYC19,HsiaoTK20} and us.\cite{note10} The second building block would be the non-adiabatic single electron pump, not previously demonstrated in dopant-free GaAs/AlGaAs.

In this Letter, we demonstrate non-adiabatic single-electron pumps in dopant-free GaAs 2DEGs suitable for quantum optoelectronics applications, with one-gate pumping, operating in zero magnetic field at temperatures of up to $T=5$ K, and using a simple RF sine waveform up to 0.95 GHz. We present data for three samples (A1, A2, and B1) using two different gate layouts, design A and design B.\cite{note17} All three samples demonstrated quantized charge pumping.

\begin{figure}[t]
    \includegraphics[width=1.0\columnwidth]{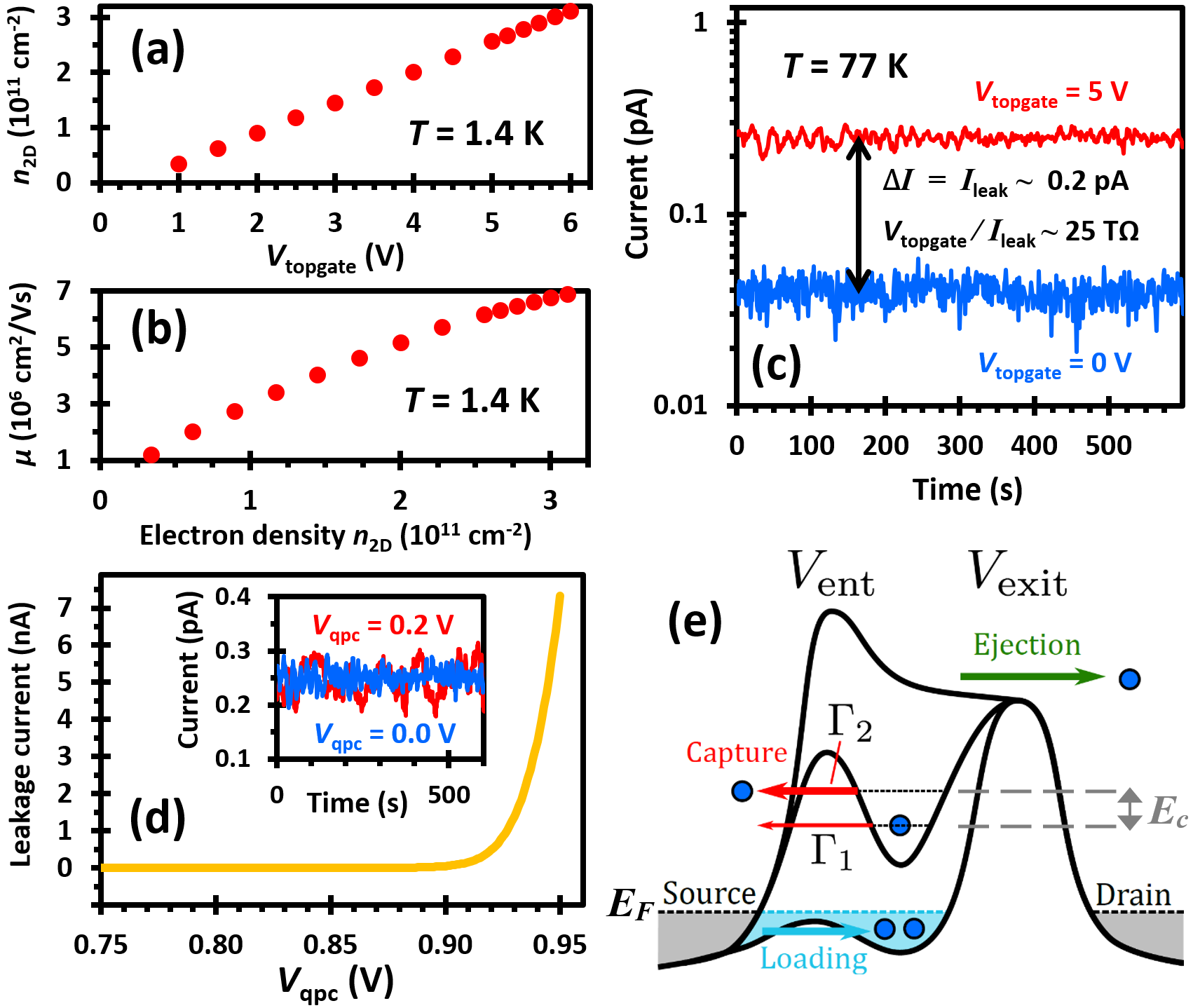}
    \caption{(a) Electron density and (b) mobility in wafer G371. (c) Gate leakage characteristics of $V_{\text{topgate}}$. (d) Gate leakage characteristics of $V_{\text{qpc}}$ at $T = 77$\,K, while $V_{\text{topgate}} = 5$\,V. At $V_{\text{qpc}}=0.2$ V, $I_{\text{leak}}$ cannot be resolved. (e) Potential energy diagram showing the pumping mechanism, in a cut along the arrow in Fig.\,\ref{Fig-SEM} at three distinct stages of a single RF cycle: loading, capture, and ejection. \textbf{Loading}. The entrance barrier $V_{\text{ent}}$ is brought below the Fermi level $E_F$ (dashed black line) and many electrons (blue circles) load into the dot from the source 2DEG. \textbf{Capture}. As the entrance barrier is raised, electrons backtunnel into the source 2DEG. Higher energy electrons have faster backtunneling rates $\Gamma_{i+1} > \Gamma_i$. A small number of electrons are captured and remain in the quantum dot as the entrance barrier is raised. \textbf{Ejection}. When the entrance barrier rises above that of the exit barrier $V_{\text{exit}}$, a fixed number of electrons are ejected from the quantum dot into the drain 2DEG. The entrance barrier is then lowered back to the loading stage in the next RF cycle.}
    \label{Fig-RFcycle}
\end{figure}  % Original fig file: Fig 2 - RF cycle v9 hi-res.png

Samples A1-A2 (B1) were fabricated on wafers G371 (G370), grown by molecular beam epitaxy with the following sequence of layers: starting from a 3" semi-insulating (SI) GaAs (100) substrate, a 200 nm GaAs buffer, a 20-period smoothing superlattice composed of a 2.5 nm GaAs layer and 2.5 nm Al$_{0.3}$Ga$_{0.7}$As layer, a 500 nm GaAs layer, a 80 nm (65 nm) Al$_{0.3}$Ga$_{0.7}$As barrier for samples A1-A2 (B1), and a 10 nm GaAs cap layer. There was no doping anywhere in the structure. Using the field effect from a topgate, a 2DEG is induced (enhancement mode) at the GaAs/AlGaAs interface located 90 nm (75 nm) below the surface for wafer G371 (G370). Transport through the single-electron pump was oriented along the high electron mobility crystal direction $[1\bar{1}0]$. Details for the fabrication of recessed ohmic contacts are otherwise identical to and extensively described in Refs.\,\onlinecite{Wendy10,ChenJCH12}. Other fabrication details are described in Figure \ref{Fig-SEM}.

\begin{figure}[t]
    \includegraphics[width=1.0\columnwidth]{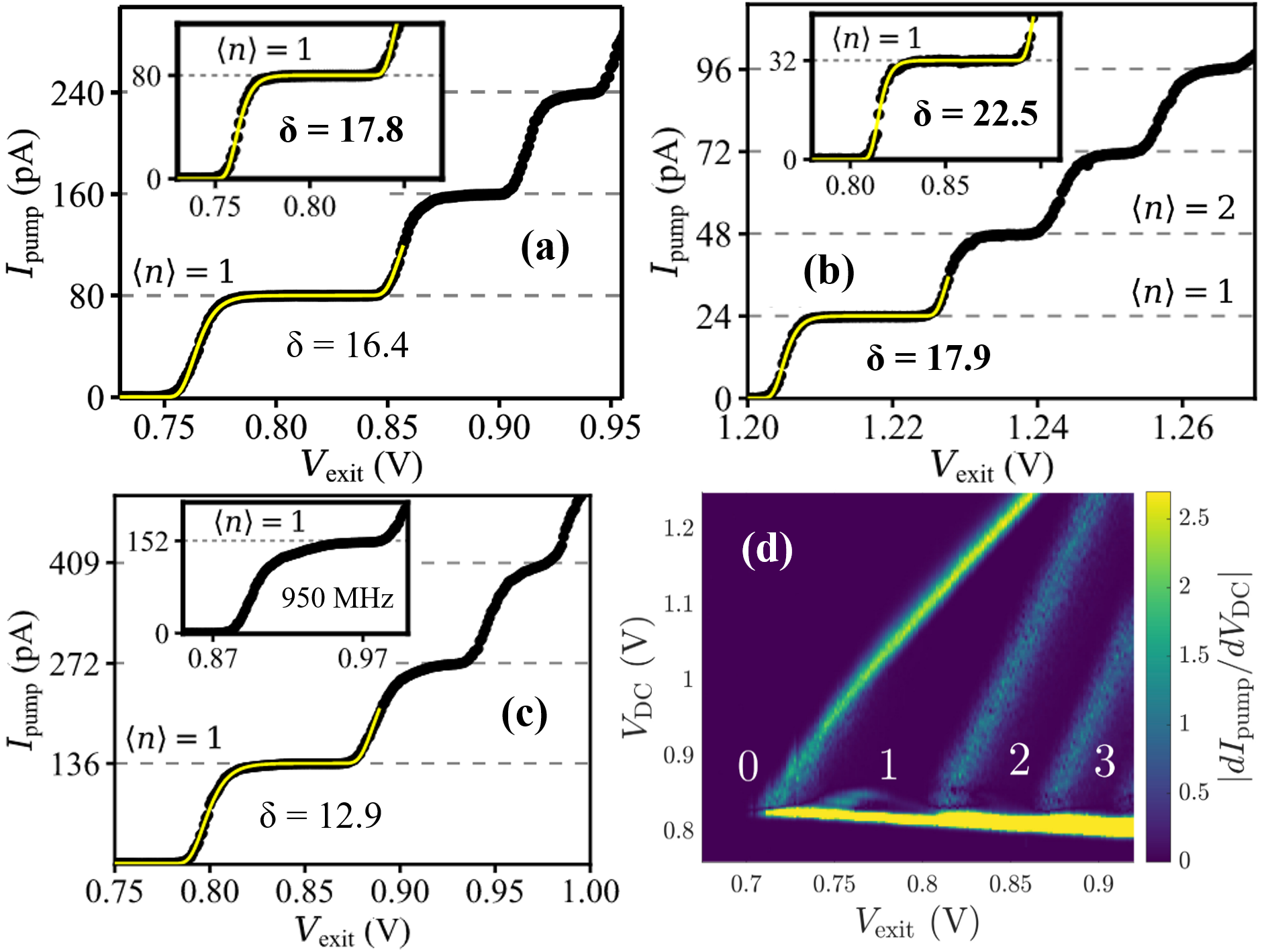}
    \caption{All data taken at $T=1.4$ K. (a\textendash c) Quantized pumped current (black dots) in samples A1, A2, and B1. The dashed horizontal lines indicate the predicted current quantization $I_{\text{pump}}=nef$, and the solid yellow lines are fits of the first quantized current plateau to eqn.\,(1). The figure of merit $\delta$ determined from the fits are stated. Insets have the same units as the main figures. Individual panels show quantized current in: (a) device A1 at $f=500$ MHz with (main) $V_{\textsc{rf}}=1.0$ V and (inset) $V_{\textsc{rf}}=1.1$ V; (b)(main) device B1 at $f=150$ MHz, (inset) device A2 at $f=200$ MHz; (c) device A1 at (main) $f=850$ MHz, and (inset) $f=950$ MHz. (d) Pump map of device A1. The colorscale represents the derivative $|dI_{\text{pump}}/dV_{\textsc{dc}}|$. Plateaus are labeled according to the number of electrons pumped per RF cycle. Additional information on experimental parameters for data shown in all panels (a\textendash d) can be found in the Supplementary Material.}
    \label{Fig-Inef}
\end{figure}  % Original fig file: Fig 3 - pump v10 hi-res.png

At $V_{\text{topgate}}=0$, the bandstructure is nearly flat since there is no intentional doping, and there are no carriers \textendash\, whether electrons or holes \textendash\, anywhere in the heterostructure. A positive voltage beyond a threshold value must be applied for a 2DEG to form. Figure \ref{Fig-RFcycle}a shows that electrons start to populate the 2DEG at $V_{\text{topgate}} = 0.4$~V (the $x$-axis intercept), and the electron density $n_{\textsc{2d}}$ varies linearly with $V_{\text{topgate}}$. At $n_{\textsc{2d}} = 3\times 10^{11}$ cm$^{-2}$, the mobility for wafer G371 is $\mu = 7\times 10^6$ cm$^2$/Vs [Fig.\,\ref{Fig-RFcycle}b]. The corresponding numbers for wafer G370 are $\mu = 2.3\times 10^6$ cm$^2$/Vs at $n_{\textsc{2d}} = 3.1\times 10^{11}$ cm$^{-2}$, and the threshold voltage for the 2DEG to start forming is $V_{\text{topgate}} = 0.5$~V.

Figure \ref{Fig-RFcycle}c shows typical small leakage current ($I_{\text{leak}}\sim 0.2$ pA) from a topgate to the 2DEG at $T=77$ K. This leakage current is expected to be even smaller at $T=1.4$ K. At $V_{\text{qpc}}=+0.2$ V, typical gate leakage to the 2DEG is negligible [Fig.\,\ref{Fig-RFcycle}d, inset].\cite{Note19} Soft breakdown occurs at $V_{\text{qpc}}\approx +0.93$V [Fig.\,\ref{Fig-RFcycle}d, main].\cite{note20} This is consistent with similar gate leakage tests on dopant-free quantum dots reported in the literature [Fig.\,2b in Ref.\,\onlinecite{Wendy13}]. The higher operating voltages without gate leakage of surface gates in dopant-free devices can be explained by the absence of doping in the $i$\nobreakdash-AlGaAs layer. The latter is in effect a better insulator than doped $n$\nobreakdash-AlGaAs in conventional 2DEGs, where surface gates leak at +0.7 V and often at much lower voltages.

Unlike all other gate voltages, $V_{\text{ent}}$ is an RF gate: $V_{\text{ent}} = V_{\textsc{dc}} + V_{\textsc{rf}}\sin(2\pi f t)$, where $V_{\textsc{dc}}$ and $V_{\textsc{rf}}$ are DC and RF voltages, respectively, combined together using a bias tee. Note the ``box'' shape of the $V_{\text{qpc}}$ gates in Fig.\,\ref{Fig-SEM}: its purpose is to specifically exclude the 2DEG from the vicinity of the $V_{\text{ent}}$ gate. This eliminates parasitic coupling between the 2DEG and $V_{\text{ent}}$, and the associated RF heating of nearby electrons.\cite{ChanKW11,Yamahata14B,MasayaPC2020} Under the combined actions of the $V_{\text{qpc}}$, $V_{\text{exit}}$, $V_{\text{ent}}$, and $V_{\text{topgate}}$ gates, a quantum dot forms and dissolves during every RF cycle [Fig.\,\ref{Fig-RFcycle}e], which is why non-adiabatic pumps are sometimes called dynamic quantum dots.\cite{WrightSam11,Waldie15,JohnsonN19}

Figure \ref{Fig-Inef} demonstrates quantized current plateaus at the expected average number of electrons pumped per RF cycle $\langle n\rangle$ for all three samples (A1, A2, B1). The universal decay cascade model\cite{Kashcheyevs10,Fricke11,Fletcher12,Kashcheyevs12,Kashcheyevs14} is fit to the quantized  $\langle n\rangle = 1$ current plateaus, with the resulting curve (yellow solid line) over the experimental data:
\begin{equation}
I = ef\sum_{j=1,2} \exp \bigg(-\exp \big[ -a(V_{\text{exit}}-V_0)+\delta (j-1) \big] \bigg) \label{eq:model}
\end{equation}
\noindent where $a$ and $\delta$ are fitting parameters, and $V_0$ is the position in gate voltage of the first plateau. The parameter $\delta$ is often used as a figure of merit. The larger the value of $\delta$, the better the current quantization (flatter plateau and steeper plateau steps) and its accuracy.\cite{Kaestner08A,Kaestner08B,Kashcheyevs10,Giblin12} Formally, $\delta  = \ln(\Gamma_2/\Gamma_1) + E_c/\Delta_{\text{ptb}}$, where $\Gamma_1$ and $\Gamma_2$ are the backtunneling rates during the capture stage of the RF cycle [see Fig.\,\ref{Fig-RFcycle}e], $E_c$ is the maximum charging energy of the dynamic quantum dot, and $\Delta_{\text{ptb}}$ is the plunger-to-barrier ratio of the entrance gate $V_{\text{ent}}$.\cite{Kashcheyevs12,Kashcheyevs14} It can be further shown that $\ln(\Gamma_2/\Gamma_1) \sim E_c/k_B T_0$, where $T_0$ is the transition temperature between the thermal hopping regime and the quantum tunneling regime.\cite{Fletcher12,Giblin19,Kaestner15,Yamahata14B} Thus $\delta$ can be expressed in terms of only the charging energy $E_c$ of the dynamic quantum dot.

The values of $\delta$ in bold obtained from the fits to eqn.\,(\ref{eq:model}) in devices A1, A2, and B1 [Figs.\,\ref{Fig-Inef}a and \ref{Fig-Inef}b] compare favorably with those from the literature in GaAs-based pumps (\textit{e.g.}, $\delta=4.6$ at 100 MHz in Fig.\,5 from Ref.\,\onlinecite{Giblin12}, $\delta=6.4$ at 500 MHz in Fig.\,3a from Ref.\,\onlinecite{Kaestner09}, $\delta=7.3$ at 600 MHz in Fig.\,2a from Ref.\,\onlinecite{Ahn17A}, and $\delta=9.6$ at 100 MHz in Fig.\,2c from Ref.\,\onlinecite{SeoM14}), \textit{in identical or closest experimental conditions}. The latter means at the same frequency $f$, in zero magnetic field, using a simple RF sinewave (not shaped RF pulses), without samples undergoing bias cooling,\cite{Pioro05-B} and using symmetric gate voltage configurations (corresponding to $V_{qpc1} = V_{qpc2}$ in our case). Indeed, large magnetic fields,\cite{Fletcher12} shaped RF waveforms,\cite{Giblin12} bias cooling,\cite{Ahn17A} and asymmetric gate voltage configurations for the dynamic quantum dot\cite{SeoM14,Ahn17A} are all experimental techniques known to significantly increase the value of $\delta$. For example, even modestly increasing magnetic field on sample A1 increased its figure of merit from $\delta=17.8$ at $B=0$ T to $\delta=21.3$ at $B=5$ T [see Fig.~S1 in the Supplementary Material]. We therefore do not compare our results to the state of the art in GaAs-based pumps for quantum metrology,\cite{Bae15,Giblin12,Stein17,Stein15,Giblin17,Bae20} since high magnetic fields and/or shaped RF pulses are routinely used. Furthermore, we do not compare our results to state of the art Si-based pumps\cite{Yamahata16,ZhaoR17,Giblin20} because p-i-n junctions in Si do not emit light, the first component of our proposed all-electrical single photon source.

\begin{figure}[t]
    \includegraphics[width=0.95\columnwidth]{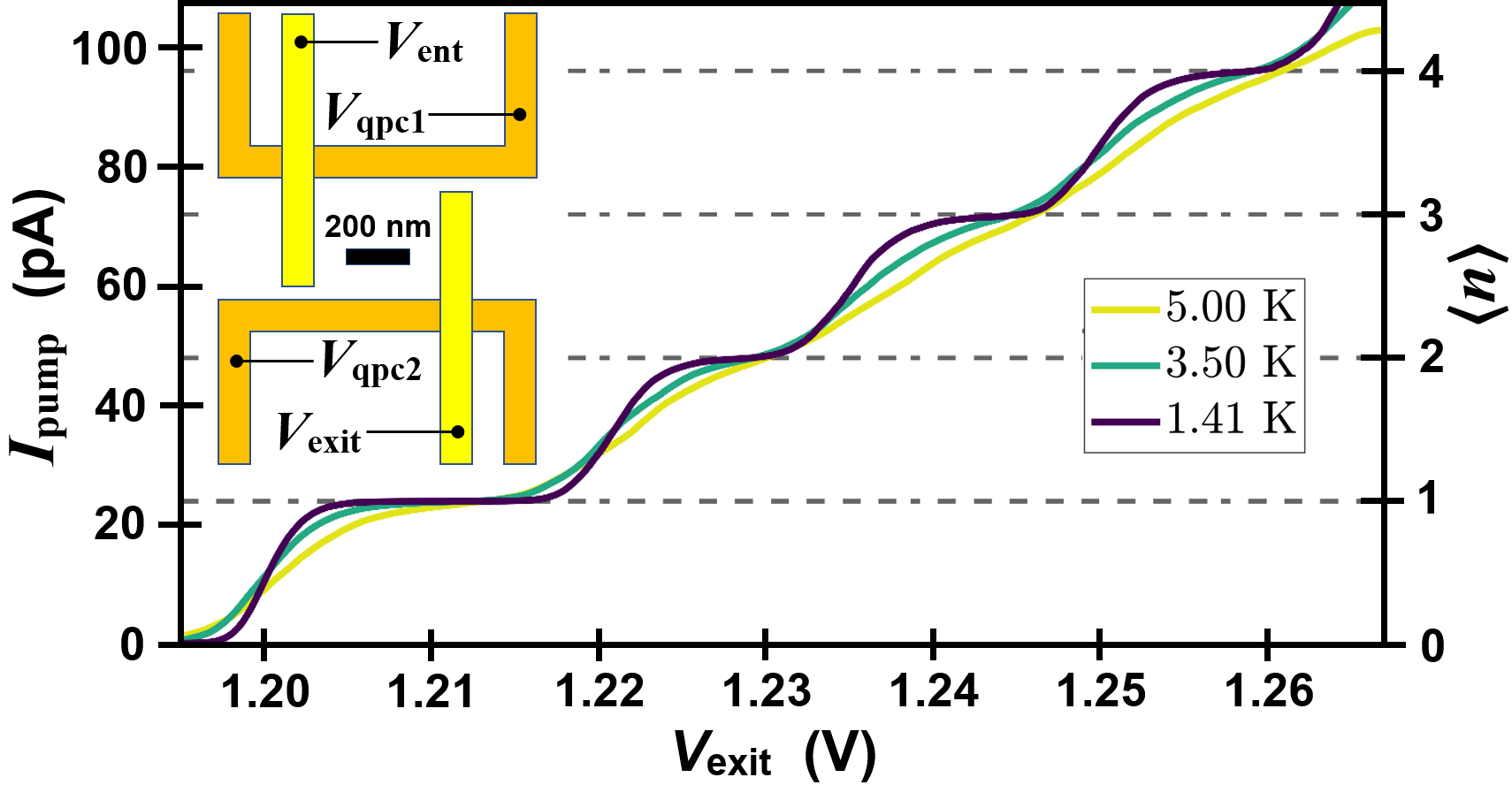}
    \caption{Temperature dependence of single-electron pumping in device B1$^\prime$ (see main text), with experimental parameters $f = 150$ MHz, $V_{\text{topgate}}$ = 5.0 V, $V_{\textsc{dc}}$ = 1.0 V, $V_{\textsc{rf}}$ = 1.2 V, and $V_{\text{qpc1}} = V_{\text{qpc2}}$ = 0.15 V. The upper inset shows a top view of the modified gate layout of device B1$^\prime$ relative to device A1.}
    \label{Fig5-SEP6}
\end{figure}  % Original fig file: Fig 5 - SEP6 v5 hi-res.png

The $\delta$ values obtained in Figure \ref{Fig-Inef} are suitable for quantum optoelectronics applications. In quantum optics, the lowest second order correlation function $g^{(2)}(\tau)=\langle I(t)I(t+\tau) \rangle / \langle I(t) \rangle^2$ at $\tau$\,=\,0 cited in the literature for single photon sources range from $g^{(2)}(0)=0.0004$ to $g^{(2)}(0)=0.003$,\cite{Somaschi16,Miyazawa16,ChenYan18,LiuJin19} corresponding to relative uncertainties in undesirable two-photon coincidences ranging from 400 ppm to 3000 ppm. To obtain similar $g^{(2)}(0)$ values in a hypothetical single photon source with an integrated single electron pump operating in the $\langle n\rangle = 1$ regime, the rate of $n=2$ pumping errors (the primary driver of two-photon emissions) would have to be 400$-$3000 ppm. These are theoretically equivalent to $8 \lesssim \delta \lesssim 10$ in the decay cascade model.\cite{note21} All the $\delta$ from samples in this paper exceed this range of values. It would thus appear that the decay cascade pumping mechanism of an integrated dopant-free single electron pump would not be a limiting factor in the performance of a single photon source.

Figure \ref{Fig-Inef}c shows pumping current at frequencies approaching 1 GHz. As expected, current quantization degrades with increasing frequency.\cite{SeoM14} Indeed, to achieve high $\delta$ values, the following conditions must be satisfied: (i) $f \ll \Gamma_L$ in the loading phase, (ii) $f \gg \Gamma_L$ in the capturing phase, and (iii) $f \ll \Gamma_R$ in the ejection phase, where $\Gamma_L$ is the backtunneling rate from the dynamic quantum dot towards the source 2DEG and $\Gamma_R$ is the tunneling rate towards the drain 2DEG. As frequency increases, at least one of these conditions, most likely the loading condition, is not met.\cite{Giblin12,Ahn17A,Yamahata14B} Nevertheless, it would appear possible for a hypothetical single photon source with an integrated single electron pump to access the $g^{(2)}(0) < 0.0004$ regime while operating at $f=850$ MHz,\cite{note23} a photon emission rate that would be competitive with state of the art single photon sources.

Figure \ref{Fig-Inef}d shows a pump map\cite{WrightSam11} of device A1. Of note, RTS events were rarely seen in these devices, and the first quantized plateau extends over a wide gate voltage range ($\sim$\,80 mV for $V_{\text{exit}}$ and more than 400 mV for $V_{\text{ent}}$, along the longest cut of either axis). For comparison, the $\langle n \rangle = 1$ plateau in state of the art GaAs-based pumps is only 20-40 mV long in exit gate voltage.\cite{Blumenthal07,Giblin12,Stein15,Stein17} This suggests dopant-free single-electron pumps could be robust against small changes in the experimental control parameters,\cite{note18} a requirement for any primary metrological standard,\cite{Stein15,Stein17,Giblin10,Giblin17}
whether optical\cite{note22} or electrical.

Figure \ref{Fig5-SEP6} shows four quantized plateaus at $T=1.4$ K in sample B1. However, this sample degraded significantly between the measurements of Fig.\,\ref{Fig-Inef}b and Fig.\,\ref{Fig5-SEP6}, such that we'll refer to this sample thereafter as B1$^\prime$. Its $\langle n\rangle = 1$ plateau now yields a smaller $\delta = 11.4$, and is transitioning from the decay cascade regime to the thermal regime.\cite{Lafarge91,Yamahata11} At $T=3.5$ K, the $\langle n\rangle = 1$ plateau is well into the thermal regime. Nonetheless, despite the lowest $\delta$ reported here and hence the smallest charging energy $E_c$ of the dynamic quantum dot, the $\langle n\rangle = 1$ plateau remains quantized at temperatures up to 5 K, easily accessible in typical optical cryostats. Since we would expect samples A1/A2/B1 (with $17.8 \leqslant \delta \leqslant 22.5$) to remain in the decay cascade regime at much higher temperatures than sample B1$^\prime$, we therefore do not believe temperatures up to 5 K prevent undoped single electron pumps from being integrated into single photon sources.

Possible applications for non-adiabatic single electron pumps extend beyond standards for current in quantum metrology\cite{Giblin12} and single photon sources in quantum optoelectronics.\cite{Chunnilall14} Examples include not only a voltage standard in quantum metrology,\cite{Hohls12} but also the fields of single electronics\cite{OnoY05} and single electron optics.\cite{Bocquillon14} In the latter, non-adiabatic single electron pumps have already made an impact on the field, with a trapping/counting scheme for hot electrons,\cite{Freise20} partitioning of on-demand electron pairs,\cite{Ubbelohde15} and on-demand emission of electron pairs with deterministically controlled exchange symmetry.\cite{Wenz19}

In conclusion, we have demonstrated non-adiabatic single electron pumps in dopant-free GaAs/AlGaAs 2DEGs operating at near-GHz frequencies, high temperature, and zero magnetic field. Given the experimental conditions, these pumps have achieved remarkable performance, and are promising as key components in optoelectronics applications such as single photon sources. If their accuracy could be further improved, they offer a possible route towards more accessible quantum standards for current, by significantly reducing the complexity of the measurement infrastructure required (dilution refrigerators, large superconducting magnets, and high frequency hardware for shaped RF pulses).

See the Supplementary Material for additional information on experimental parameters for all data shown in Figure \ref{Fig-Inef}, and for plots showing the improvement of $\delta$ in magnetic field for sample A1.

The authors thank Christine Nicoll and Masaya Kataoka for useful discussions. B.B., F.S., and A.S. contributed equally to this paper. S.R.H. acknowledges support from a Waterloo Institute for Nanotechnology (WIN) Nanofellowship. This research was undertaken thanks in part to funding from the Canada First Research Excellence Fund (Transformative Quantum Technologies), Defence Research and Development Canada (DRDC), and the Natural Sciences and Engineering Research Council (NSERC) of Canada. The University of Waterloo's QNFCF facility was used for this work. This infrastructure would not be possible without the significant contributions of CFREF-TQT, CFI, ISED, the Ontario Ministry of Research and Innovation, and Mike and Ophelia Lazaridis. Their support is gratefully acknowledged. F.H. has received funding from the EMPIR programme co-financed by the Participating States and from the European Union’s Horizon 2020 research and innovation programme, grant 17FUN04 SEQUOIA, and by the Deutsche Forschungsgemeinschaft (DFG, German Research Foundation) under Germany’s Excellence Strategy – EXC-2123 QuantumFrontiers – 390837967.

The data that support the findings of this study are available from the corresponding author upon reasonable request.


\begin{thebibliography}{99}
\bibitem[{not({\natexlab{a}})}]{note10}
\bibinfo{note}{Details of our lateral p-i-n junctions will be extensively
  covered in another publication, and will not be described here.}
\bibitem[{not({\natexlab{b}})}]{note17}
\bibinfo{note}{A comparative study on the effects of gate layout on pump
  performance in dopant-free GaAs devices will be addressed in another
  publication.}
\bibitem[{not({\natexlab{a}})}]{note20}
\bibinfo{note}{$V_{\text{qpc}}=[0.9,1.0]$ V is the typical soft breakdown
  voltage range.}
\bibitem[{not({\natexlab{b}})}]{note21}
\bibinfo{note}{At the optimal operating point of a single electron pump in the
  $\langle n\rangle = 1$ regime, where the uncertainty in pumped current is the
  lowest, the rates for $n=0$ and $n=2$ pumping errors are
  equal.\cite{Kaestner15,Kashcheyevs10,Fricke13,Giblin16} At this optimal
  operating point, a lower bound for the pumped current uncertainty can be
  estimated from the experimental $\delta$
  value,\cite{Kashcheyevs10,Giblin12,SeoM14,ZhaoR17} using eqn.\,(3.10) from
  Ref.\,\onlinecite{Kaestner15}: $P_{\text{error}} = 1 - P_1 \approx 2\delta
  \exp(-\delta)$.}
\bibitem[{not({\natexlab{c}})}]{note23}
\bibinfo{note}{Because $\delta_{f=850\text{MHz}} > \delta_{400
  \text{~ppm~for~}n=2\text{~pumping~errors}}$ or, more explicitly, $12.9>10$.}
\bibitem[{not({\natexlab{d}})}]{note18}
\bibinfo{note}{In principle, a poor lever arm for {$V_{\text{ent}}$} and
  {$V_{\text{exit}}$} could account for the large voltage range of the current
  plateaus. However, the gates in device A1 displayed large lever arms.
  Generally, we have observed that samples with poor gate lever arms tend to
  produce low-quality (small {$\delta$}) current quantization, which is
  certainly not the case for device A1.}
\bibitem[{not({\natexlab{e}})}]{note22}
\bibinfo{note}{A high-rate, high-fidelity single photon source could be used as
  a metrological-grade standard for photon intensity at emission wavelengths
  other than 555 nm (the SI unit candela is defined at $\lambda \approx 555$
  nm).}
\end{thebibliography}
\end{document}

\title{ SUPPLEMENTARY MATERIAL: \\
Non-adiabatic single-electron pumps in a dopant-free GaAs/AlGaAs 2DEG}

\author{B. Buonacorsi}
\affiliation{Institute for Quantum Computing, University of Waterloo, Canada}
\affiliation{Department of Physics, University of Waterloo, Canada}
\author{F. Sfigakis}
\affiliation{Institute for Quantum Computing, University of Waterloo, Canada}
\affiliation{Northern Quantum Lights Inc., Canada}
\affiliation{Department of Chemistry, University of Waterloo, Canada}
\author{A. Shetty}
\affiliation{Institute for Quantum Computing, University of Waterloo, Canada}
\affiliation{Department of Chemistry, University of Waterloo, Canada}
\author{M. C. Tam}
\author{H. S. Kim}
\affiliation{Dept.~of Electrical and Computer Engineering, University of Waterloo, Canada}
\affiliation{Waterloo Institute for Nanotechnology, University of Waterloo, Canada}
\author{S. R. Harrigan}
\affiliation{Institute for Quantum Computing, University of Waterloo, Canada}
\affiliation{Department of Physics, University of Waterloo, Canada}
\affiliation{Waterloo Institute for Nanotechnology, University of Waterloo, Canada}
\author{\\ F. Hohls}
\affiliation{Physikalisch-Technische Bundesanstalt (PTB), Germany}
\author{M. E. Reimer}
\affiliation{Institute for Quantum Computing, University of Waterloo, Canada}
\affiliation{Department of Physics, University of Waterloo, Canada}
\affiliation{Northern Quantum Lights Inc., Canada}
\affiliation{Dept.~of Electrical and Computer Engineering, University of Waterloo, Canada}
\author{Z. R. Wasilewski}
\affiliation{Institute for Quantum Computing, University of Waterloo, Canada}
\affiliation{Department of Physics, University of Waterloo, Canada}
\affiliation{Northern Quantum Lights Inc., Canada}
\affiliation{Dept.~of Electrical and Computer Engineering, University of Waterloo, Canada}
\affiliation{Waterloo Institute for Nanotechnology, University of Waterloo, Canada}
\author{J. Baugh}
\affiliation{Institute for Quantum Computing, University of Waterloo, Canada}
\affiliation{Department of Physics, University of Waterloo, Canada}
\affiliation{Northern Quantum Lights Inc., Canada}
\affiliation{Department of Chemistry, University of Waterloo, Canada}
\affiliation{Waterloo Institute for Nanotechnology, University of Waterloo, Canada}

\begin{abstract}
\end{abstract}

\maketitle

\begin{table}[h]
    \begin{ruledtabular}
    \begin{tabular}{ccccccc}
    Panel& Sample & $f$ 
    & $V_{\textsc{rf}}$ & $V_{\textsc{dc}}$ 
    & $V_{\text{qpc}}$ & $V_{\text{topgate}}$  \\
    in Fig.\,3 & ID & (MHz) & (V) & (V) & (V) & (V) \\ \hline
    (a) main & A1 & 500 & 1.00 & 0.87 & 0.18 & 6.0 \\
    (a) inset & A1 & 500 & 1.10 &  0.87 & 0.18 & 6.0 \\
    (b) main & B1 & 150 & 0.80 & 1.0 & 0.15 & 5.0 \\
    (b) inset & A2 & 200 & 0.80 & 0.71 & 0.20 & 5.5 \\
    (c) main & A1 & 850 & 1.60 & 0.94 & 0.18 & 6.0 \\
    (c) inset & A1 & 950 & 1.75 & 0.90 & 0.18 & 6.0 \\
    (d) map & A1 & 500 & 1.10 & $var.$ & 0.18 & 6.0 \\    
    \end{tabular}
    \end{ruledtabular}
    \caption{Additional information for experimental parameters for all panels in Figure 3 of the main text. Note that $V_{\text{qpc}}=V_{\text{qpc1}}=V_{\text{qpc2}}$. For the pump map shown in panel (d), the numerical derivative $|dI_{\text{pump}}/dV_{\textsc{dc}}|$ was performed after a 3-point moving average. A line cut along the $V_{\text{exit}}$ direction would appear similar to the traces in panels (a)$-$(c).}
    \label{TableS1}
\end{table}

~\\~\\~\\~\\~\\~\\~\\~\\

\begin{figure}[t]
    \includegraphics[width=1.0\columnwidth]{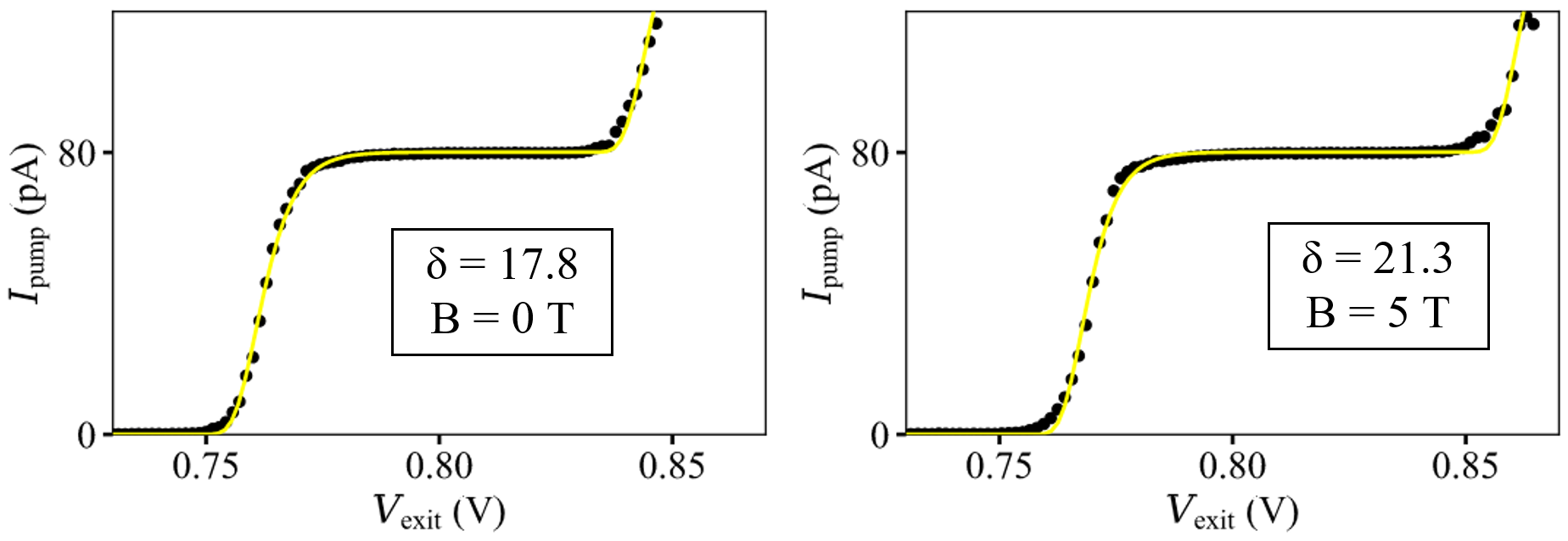}
    \caption{The two panels above show the improvement of $\delta$ with magnetic field in sample A1 at $T = 1.4$\,K with $f=500$\,MHz, $V_{\text{topgate}}=6.0$\,V, $V_{\textsc{dc}}=0.87$\,V, $V_{\textsc{rf}}=1.1$\,V, and $V_{\text{qpc1}}=V_{\text{qpc2}}=0.18$\,V. The black circles are the experimental data points, and the yellow solid lines are the fits to eqn.(1) in the main text. The resulting $\delta$ from the fits are stated in the boxed labels.}
    \label{FigBla}
\end{figure}  % Original fig file: Fig 5 - SEP6 v5 hi-res.png

\clearpage